\documentstyle[sprocl]{article}

\input{psfig}

\begin{document}

\title{Monte Carlo simulation of systems with complex-valued measures}

\author{J.F. Markham and T. D. Kieu}
\address{School of Physics, University of Melbourne,\\ Vic 3052, Australia}

\maketitle\abstracts{
A simulation method based on the RG blocking is shown to yield 
statistical errors smaller than that of the crude MC using absolute values 
of the original measures.  The new method is particularly suitable to apply
to the sign problem of indefinite or complex-valued measures.  We
demonstrate the many advantages of this method in the simulation of
2D Ising model with complex-valued temperature.
}

\section{The Sign Problem}
In order to evaluate a multi-dimensional integral of the partition
function
\begin{eqnarray}
Z&\equiv& \int fdV
\label{1a}
\end{eqnarray}
using Monte Carlo (MC) one can sample the points in the integration
domain
with a non-uniform distribution, $p$.
This sampling gives the following estimate for the integral:
\begin{eqnarray}
Z &\approx& \left\langle
{f}/{p}
\right\rangle \pm \sqrt{{S}/{N} },
\label{1}
\end{eqnarray}
where $N$ is the number of points sampled, $p\geq 0$ and is normalised;
\begin{eqnarray}
\left\langle f/p \right\rangle&\equiv&\frac{1}{N}
\sum\limits_{i=1}^{N}f(x_{i} )/p(x_{i}).\nonumber
\end{eqnarray}
and
\begin{eqnarray}
S&\equiv& \int\left|{f}/{p}-Z\right|^2 p dV, \nonumber\\
&\approx& \left\langle {f^{2} }/{p^{2} }
\right\rangle -\left\langle {f}/{p} \right\rangle ^{2}.
\label{2}
\end{eqnarray}

The best choice of $p$ is the one that minimises the standard
deviation squared $S$.  This can be found
by variational method leading to the \underline {crude average-sign
MC weight}
\begin{eqnarray}
p_{\rm crude}&=&{\left| f\right| }\left/{\int \left| f\right| dV }\right.,
\label{3}
\end{eqnarray}
giving the optimal
\begin{eqnarray}
S_{\rm crude} &=&\left( \int \left| f\right| dV \right) ^{2} -\left|
\left( \int fdV \right) \right| ^{2}.
\label{4}
\end{eqnarray}

The \underline {sign problem}~\cite{sign} arises when
$Z/\int|f|dV$ is vanishingly small: then unless a huge number of
configurations are MC sampled, the large statistical fluctuations
of the partition function render the measurement
meaningless.
\section{Improved Simulation Method}

One way of smoothing out the sign problem is to do part of the integral
analytically, and the remainder using MC.  The analytical summation
is not just directly over a subset of the dynamical variables; in
general it can be a renormalisation group (RG) blocking where
coarse-grained
variables are introduced.
This does yield certain improvement over the crude MC in general.

Let $P\{V',V\}$ be the normalised RG weight relating the original
variables
$V$ to the blocked variables $V'$,
\begin{eqnarray}
P\{V',V\} &\geq& 0,\nonumber\\
\int P\{V',V\} dV' &=& 1.
\nonumber
\end{eqnarray}
Inserting this unity resolution into the integral~(\ref{1a})
\begin{eqnarray}
I &=& \int dV\int dV' P\{V',V\}f,\nonumber\\
&\equiv& \int dV' g(V').
\label{5}
\end{eqnarray}
Thus, an MC estimator is only needed for the remaining integration over
$V'$
in~(\ref{5}).  As with the crude method of the last section, variational
minimisation for $S$ of~(\ref{2}), with $g$ in place of $f$, leads to
the
\underline {improved} MC
\begin{eqnarray}
p_{\rm improved}&=&{|g|}\left/{\int |g| dV'}\right. .
\label{improved}
\end{eqnarray}

Firstly, the improved
weight sampling yields a partition function of magnitude not
less than that sampled by the crude weight:
\begin{eqnarray}
\left|\left\langle \left\langle \rm sign \right\rangle
\right\rangle_{\rm improved}\right| &\equiv& {Z}\left/{\int |g| dV'}
\right.,
\nonumber\\
&=& \frac{Z}{\int\left|\int P\{V',V\}fdV\right|dV'}, \nonumber\\
&\geq& {Z}\left/{\int|f|dV}\right. ,\nonumber\\
&\equiv&
\left|\left\langle \left\langle \rm sign \right\rangle
\right\rangle_{\rm crude}\right|.
\label{i1}
\end{eqnarray}

Secondly, it is also not difficult to see that the statistical
fluctuations
associated with improved MC is not more than that of the crude MC,
\begin{eqnarray}
&&S_{\rm improved} - S_{\rm crude} \nonumber\\
&&= \int \left| \frac{g^2}{p_{\rm
improved}}
\right|dV' - \int \left| \frac{f^2}{p_{\rm crude}}\right|dV,\nonumber\\
&&= \left( \int |g| dV'\right)^2 - \left( \int |f| dV\right)^2,
\nonumber\\
&&\leq 0,
\label{i2}
\end{eqnarray}

Thus the RG blocking always reduces the
statistical fluctuations of
an observable measurement by reducing the magnitude of
${\sqrt{S}}/{\left|\left\langle \left\langle \rm sign
\right\rangle
\right\rangle\right|}$.

Note that the special
case of equality in~(\ref{i1},\ref{i2}) occurs \underline {iff} there
was no sign
problem to begin with.
How much improvement one can get out of the new MC weight
depends on the details of the RG blocking and on the original measure $f$.
\section{Complex 2D Ising Model}

The Hamiltonian for the Ising model on a square lattice is
\begin{eqnarray}
H&=&-j\sum\limits_{\left\langle nn\prime \right\rangle }s_{n} s_{n\prime
}
-h\sum\limits_{n}s_{n}.
\label{11}
\end{eqnarray}
Here we allow $j$ and $h$ to take on complex values in general.
For the finite lattice, periodic boundary conditions are used.

The phase boundaries for the complex temperature 2D Ising model
with $h=0$ are depicted in Figure 1~\cite{shrock}.

\begin{figure}
\hspace*{0.5in}
\psfig{figure=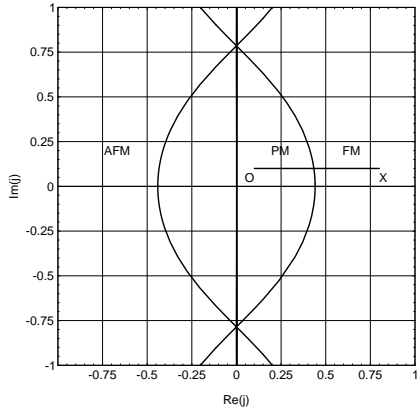,height=3.0in}
\caption{\label{Figure 1} Phase diagram in the complex $j$ plane.}
\end{figure}

For the improved method,
we adopt a simple RG blocking over the odd sites, labeled $\circ$.
That is,
the analytic summation is done over the configuration space spanned
by the even $\circ$ sites; while MC is used to evaluate the sum over the
remaining
lattice of the odd $\bullet$ sites.
Figure 2 shows the two 
sublattices, and 
how the $\bullet$ sites are to be labelled relative to the $\circ$ sites, 
for the site labelled $x$.
\begin{figure}
\hspace*{1in}
\psfig{figure=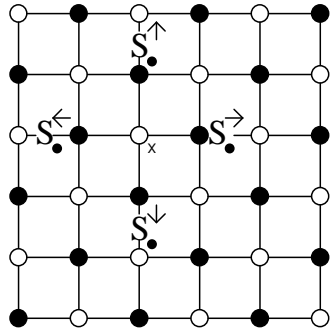,height=2.5in}
\caption{\label{Figure 2} Relative spin positions on a partitioned lattice.}
\end{figure}

In general, with finite-range interactions between the spins, one can
always subdivide the lattice into sublattices, on each of which the
spins
are (nearly) independent for the partial sum to be carried out.

Summing over the spins $s_\circ$,
\begin{eqnarray}
Z
&=&\sum\limits_{\{s_\bullet\}} e^{h\sum\limits_{\bullet sites}
s_{\bullet } }
\prod\limits_{\circ sites}2\cosh \left[ js_{\bullet }^{+} +h\right]
\end{eqnarray}
where $s_{\bullet }^{+}$ is the sum of the spins of the four odd neighbours
surrounding each even site.
The improved MC weight is then the absolute value of the summand on
the right hand side of the last expression for $Z$.

The quantities measured are magnetisation, $M$, and
susceptibility,
$\chi$.
These can be expressed in terms of the first and second derivatives of
$Z$
respectively,
evaluated at $h=0$.  
In all the simulations, square two-dimensional lattices of various sizes
with periodic boundary
conditions are used. The heat-bath algorithm
is used.
Two additional benefits arise from the improved method.
The first is that the number of sites to be visited is halved.
While the expressions to be calculated at the remaining sites turn out
to be far more complicated, the use of table look-up means that
evaluating them need not be computationally more expensive.
The second benefit is that the number of sweeps required to decorrelate
data points is reduced.

\begin{figure}
\hspace*{0.8in}
\psfig{figure=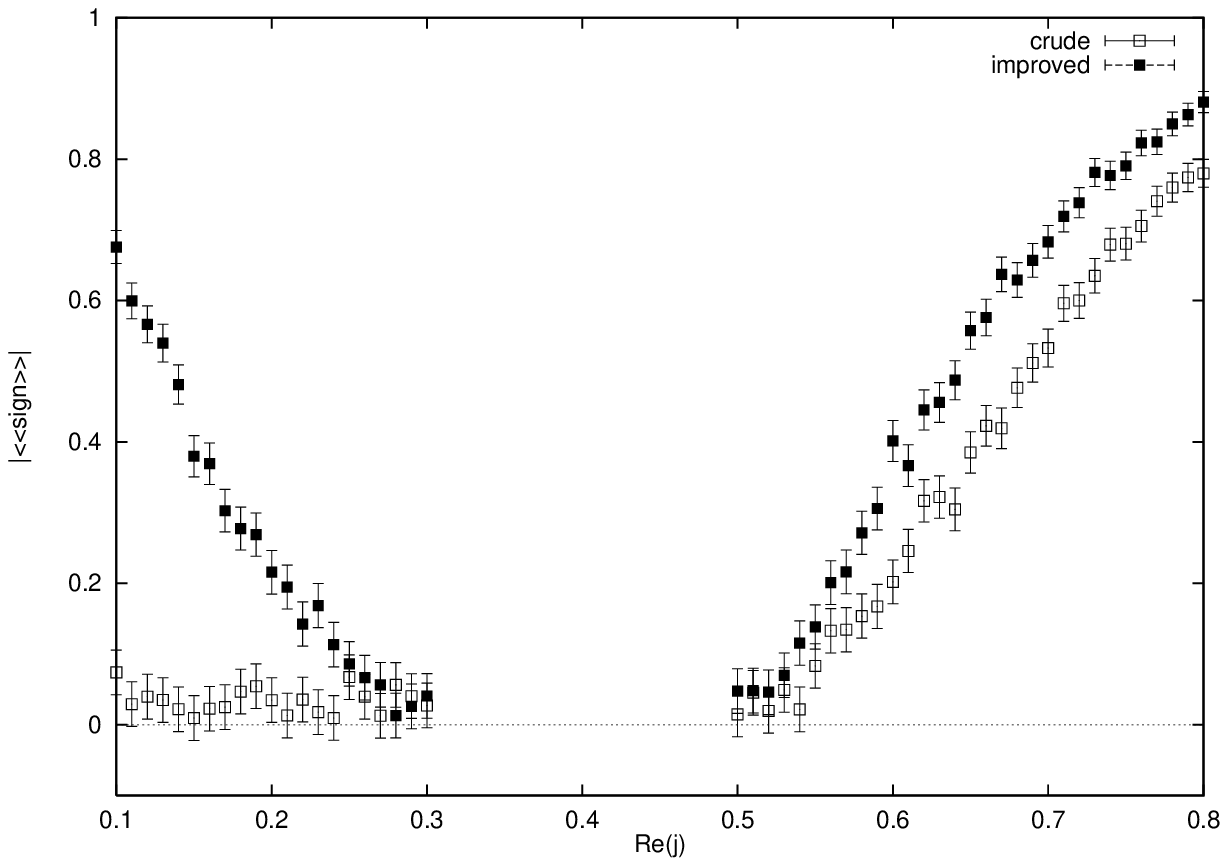,height=2.1in}
\caption{\label{Figure 3} $|<<{\rm sign}>>|$ vs $Re(j)$ $Im(j)=0$, $h=0$.
Filled boxes are the improved data.}
\end{figure}

As a test of the improved method, it is compared to the crude
one along the path $ OX$ in Figure 1.
Comparison with series-expansion data are plotted in Figure 4, in which
the discs are the absolute-value circles of complex-valued statistical 
errors for simulated results.

\begin{figure}
\vspace{-1.0in}
\hspace*{0.3in}
\psfig{figure=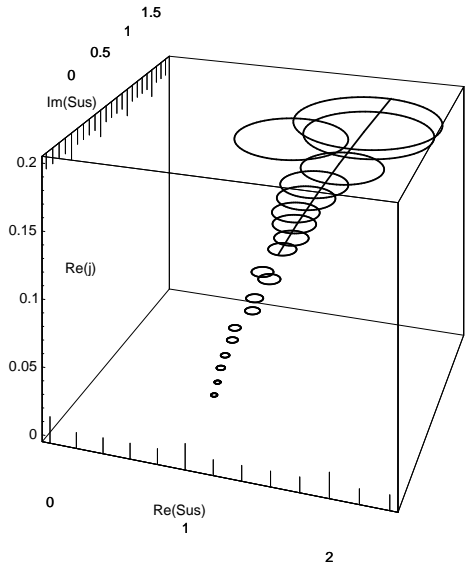,height=3.5in}
\caption{\label{Figure 4} Improved $\chi$ vs $Re(j)$; expansion data
shown as line.}
\end{figure}

\section{Concluding Remarks}
The sign problem is partially eleviated in the improved method~\cite{full}
because of some partial phase cancellation among the
original indefinite or complex-valued measure after an exact RG
transformation.
A particular RG blocking is chosen for our
illustrative example of the 2D Ising model with complex-valued measure.
But other choices of RG blocking are feasible and how effective they
are depends on the physics of the problems.
\section*{Acknowledgments}
We are indebted to Robert Shrock and Andy Rawlinson for help and discussions.
We also acknowledge the Australian 
Research Council and Fulbright Program for financial support.

\end{document}